\documentclass[10pt,conference]{IEEEtran}

\usepackage{cite}
\usepackage{amsmath,amssymb,amsfonts}
\usepackage{algorithmic}
\usepackage{graphicx}
\usepackage{textcomp}
\usepackage{xcolor}
\usepackage{tabularx}
\usepackage{booktabs}
\usepackage{enumitem}
\usepackage{xcolor}
\usepackage{flushend}
\usepackage{marvosym}

\newcommand{\ljy}[1]{\textcolor{black}{#1}}
\newcommand{\toolname}{VeriSynth}

\begin{document}

\title{Towards Automated Formal Verification of zkEVMs Using LLM-Guided Constraint Synthesis}

\author{
\IEEEauthorblockN{Shichen Huang\IEEEauthorrefmark{1}, Zhenghe Jiang\IEEEauthorrefmark{1}, Yi Jiang\IEEEauthorrefmark{2},
Ling-I Wu\IEEEauthorrefmark{3}, Jingyang Li\IEEEauthorrefmark{3}
Guoqiang Li\IEEEauthorrefmark{3}\textsuperscript{\Letter}}

\IEEEauthorblockA{\IEEEauthorrefmark{1}Shanghai Polytechnic University
}
\IEEEauthorblockA{\IEEEauthorrefmark{2}Nanjing University of Science and Technology}
\IEEEauthorblockA{\IEEEauthorrefmark{3}Shanghai Jiao Tong University
}
}
\maketitle

\begin{abstract}
\ljy{Zero-Knowledge Ethereum Virtual Machines (zkEVMs) secure Ethereum rollups by generating zero-knowledge proofs that guarantee off-chain execution correctness. However, subtle implementation bugs (e.g., incorrect gas accounting) can lead to valid proofs certifying semantically faulty states, thereby silently defeating cryptographic guarantees. Formal verification via SMT solvers can prevent this, but is bottlenecked by specification: current zkEVM development practice lacks automated methods to translate Rust opcode handlers into verification models. Current practices rely on unsustainable manual specifications, while LLM-based approaches suffer from hallucination and lack formal guarantees. To address this, we propose VeriSynth, a framework that synthesizes executable Python/Z3 verification models from Rust zkEVM code. VeriSynth enforces a hybrid paradigm: an LLM acts strictly as a formalization frontend to translate code into symbolic constraints, while an SMT solver serves as the correctness arbiter. To handle complex multi-component state transitions, VeriSynth integrates semantic decomposition, retrieval-grounded prompting, and verification-guided auto-repair into a closed-loop pipeline. We evaluate VeriSynth on the first source-level zkEVM verification benchmark, encompassing both correct and faulty opcode implementations. VeriSynth achieves a bug detection rate of over 90\%, substantially outperforming direct and conversational LLM baselines, as well as a production-grade handwritten mutation-testing suite. Ablation studies confirm that each pipeline component is critical to the framework's overall effectiveness.
}
\end{abstract}


\begin{figure*}[t]
  \centering
  \includegraphics[width=0.8\textwidth]{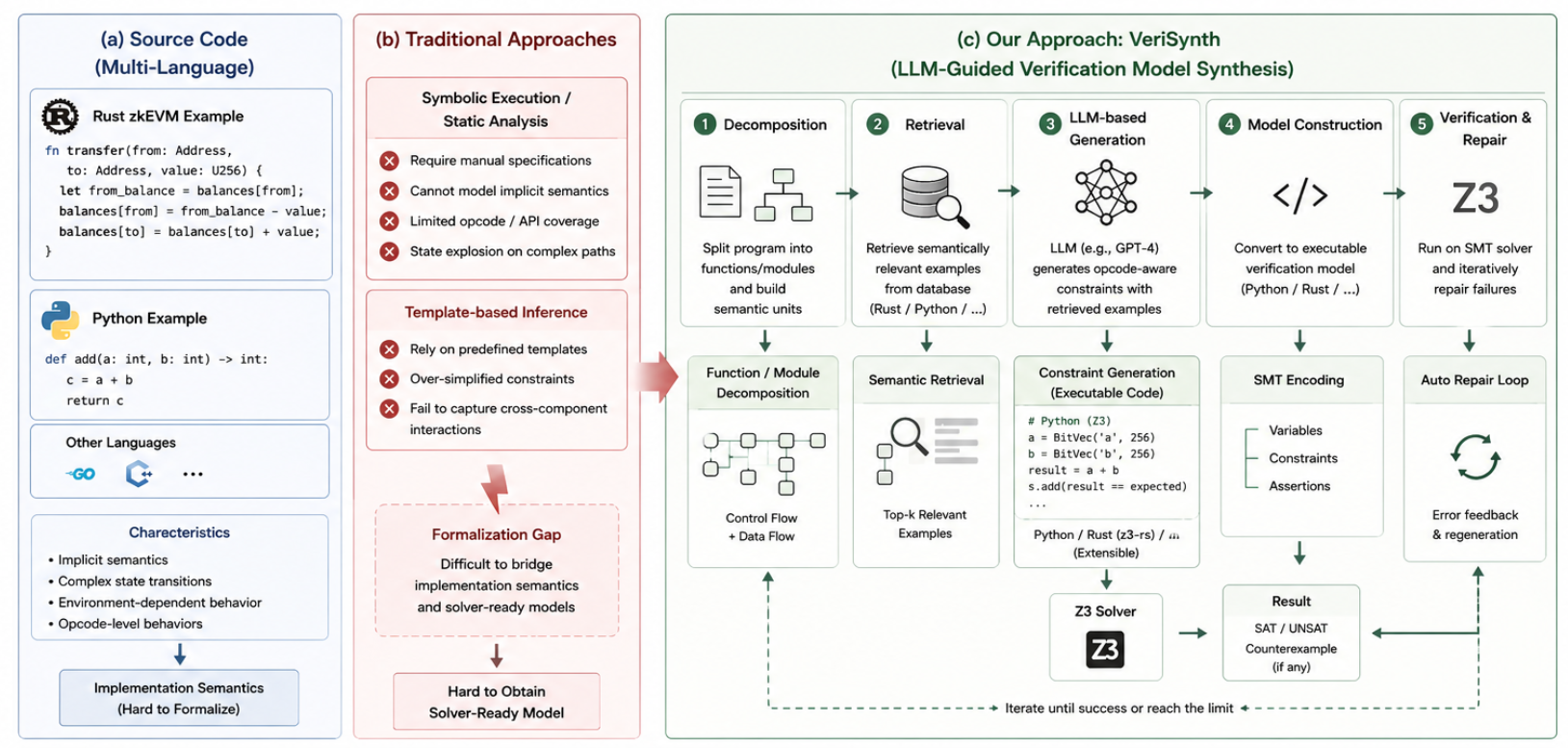}
  \caption{
  Motivation of LLM-guided verification model synthesis. 
  Existing approaches fail to translate Rust implementations into formal verification models due to implicit semantics and complex opcode behaviors. 
  Our approach bridges this gap by synthesizing executable SMT constraints that can be directly validated by a solver.
  }
  \label{fig:motivation}
\end{figure*}

\section{Introduction}

\ljy{The massive growth of decentralized software systems has led Ethereum to increasingly offload execution to Layer-2 infrastructures called rollups, which scale throughput by processing transactions off-chain~\cite{ethereumZKRollups}.
To ensure the integrity of these decoupled states, the underlying blockchain employs a Zero-Knowledge Ethereum Virtual Machine (zkEVM) to construct mathematical validity proofs that verify off-chain transitions~\cite{wood2014ethereum}. 
However, this verification pipeline relies on a fundamental software engineering assumption: the zkEVM implementation must faithfully realize the intended virtual machine semantics. 
If this implementation contains subtle bugs—such as incorrect gas accounting, missing state updates, or wrong boundary checks—the system can generate valid cryptographic proofs for semantically faulty execution states. 
These silent semantic errors bypass proof verification, creating a critical need for automated, solver-backed formal verification of the source-level zkEVM code~\cite{luu2016making,tsankov2018securify,kalra2018zeus,hildenbrandt2018kevm}.}

\ljy{Addressing this challenge involves two distinct layers: automated reasoning and verification model synthesis. 
At the reasoning layer, modern SMT solvers and symbolic reasoning engines already provide mature backends for checking logical constraints over bit-vectors, arrays, and state-transition relations~\cite{demoura2008z3,barrett2010smtlib,cadar2008klee,kroening2004cbmc}.
At the synthesis layer, however, the situation is different: solvers can only reason about the constraints they are given; they do not automatically infer specifications or verification conditions from an implementation~\cite{king1976symbolic,ernst2001daikon,flanagan2001houdini}.
This semantic gap is especially pronounced in zkEVM implementations, where complex execution behaviors are implicit in low-level Rust source code. 
Consequently, the central bottleneck in zkEVM verification is not the capability of the reasoning backend, but the lack of automated, practical tooling for translating implicit, source-level Rust semantics into accurate, solver-ready verification models at the opcode level.} 

\ljy{Current software engineering workflows do not close this synthesis gap, because they leave the constraint-modeling burden on human developers.
Leading zkEVM projects such as Scroll rely on manually written specifications and handcrafted mutation tests to detect semantic bugs~\cite{tolmach2020survey}.
However, maintaining these manual artifacts does not scale as low-level implementations undergo frequent optimization, and tests are bounded by the behaviors that developers explicitly anticipate.
Furthermore, while the community has constructed benchmarks for smart contract and EVM bytecode vulnerabilities~\cite{salzano2026empirical}, there is no standardized, source-level bug benchmark for zkEVM Rust implementations.
As a result, developers shoulder the modeling effort, and researchers lack a concrete baseline for evaluating automated verification model synthesis at the infrastructure layer.
} 

\ljy{LLM-based methods offer a potential path to reduce this manual burden, but they are not reliable enough on their own.
Existing LLM-based specification generators primarily produce high-level declarative properties~\cite{ma2025specgen,liu2024propertygpt}, whereas verifying zkEVM opcode transitions requires detailed executable models that capture multi-component state updates across stack, memory, storage, gas, and execution context~\cite{brown2020language,white2023prompt}.
More critically, direct LLM bug-finding suffers from code hallucination and provides no formal guarantees.
To scale verification to complex systems, a hybrid paradigm is needed: the LLM acts strictly as a formalization frontend that translates code into candidate executable SMT constraints, while an SMT solver serves as the correctness arbiter.
No existing automated framework realizes this paradigm for source-level Rust zkEVM implementations.}

\ljy{To address these gaps, we introduce VeriSynth, a framework that synthesizes executable
Python/Z3 verification models from zkEVM Rust source code.
Its core design separates responsibilities: an LLM serves as a formalization frontend
that translates source-level code into candidate symbolic constraints, while an SMT
solver acts as the correctness arbiter: the LLM proposes, the solver decides.
VeriSynth targets the full zkEVM execution state, guiding the LLM to extract semantics
that are otherwise implicit in Rust control flow and encoding them as symbolic variables and constraints covering stack manipulations, memory layouts, persistent storage updates, dynamic gas and refund accounting, and execution context registers.
To make these interacting components tractable, the framework integrates three mechanisms into a closed-loop pipeline: semantic decomposition isolates state transitions into modular verification units; retrieval-grounded prompting anchors variable types and solver APIs against previously validated examples; and verification-guided auto-repair iteratively refines generated models using compiler diagnostics and solver counterexamples until they compile and verify.
}

\ljy{We evaluate VeriSynth on our self-constructed source-level zkEVM verification
benchmark of 95 injected faulty samples across five opcode semantic families.
VeriSynth detects 87 out of 95 bugs, achieving a detection rate of 91.6\%,
substantially outperforming the LLM-only baseline (46.3\%) and the conversational
LLM baseline (61.1\%).
The framework remains effective across diverse opcode families, with per-family
detection rates ranging from 95.5\% (arithmetic/bitwise) to 85.7\% (call/create
paths).
Ablation confirms that auto-repair is the single largest contributor: removing it
drops the detection rate from 91.6\% to 66.3\%, as many initially generated models
fail on local syntax or sort errors rather than semantic misunderstandings.
Retrieval grounding and semantic decomposition each provide further gains.}

\ljy{To demonstrate practical engineering value, we compare VeriSynth directly with
Scroll's handwritten Rust mutation-test workflow.
Scroll's production-grade tests expose 55 out of 95 injected bugs, while
VeriSynth detects 87---32 additional bugs that escaped manually written assertions.
This gap is consistent across both semantic errors and execution-path errors,
indicating that automated constraint synthesis complements, rather than replaces,
manual testing pipelines.}

The main contributions of this paper are as follows:

\begin{itemize}

\item We propose VeriSynth, an LLM-guided verification framework that synthesizes executable Python/Z3 models directly from zkEVM Rust source. VeriSynth treats LLMs as formalization frontends and SMT solvers as final validators, integrating semantic decomposition, retrieval-grounded prompting, and verification-guided auto-repair into a closed-loop pipeline.

\item We construct the first source-level benchmark for Rust opcode implementation verification in zkEVM systems, spanning five semantic families of injection faults in control flow, state updates, and resource accounting. This benchmark targets implementation-level faults, a category distinct from existing contract- and bytecode-level datasets.

\item We conduct a systematic empirical evaluation showing VeriSynth substantially outperforms direct LLM baselines and industrial mutation-testing workflows. Ablation studies confirm that auto-repair and retrieval grounding are critical for solver compatibility and model executability.

\end{itemize}

\section{preliminary}

\subsection{Zero-Knowledge Ethereum Virtual Machines}

A zero-knowledge Ethereum Virtual Machine (zkEVM) executes Ethereum-compatible programs and generates zero-knowledge proofs showing that the resulting state transitions satisfy a predefined constraint system. In rollup systems, transactions are executed off-chain, while the corresponding proofs are verified on-chain, allowing a large batch of executions to be validated without re-executing every instruction.

An EVM execution state can be abstracted as
\[
\sigma=\langle pc, stk, mem, sto, gas, env\rangle,
\]
where \(pc\) denotes the program counter, \(stk\) the operand stack, \(mem\) the transient memory, \(sto\) the persistent storage, \(gas\) the remaining gas, and \(env\) the execution context. Each opcode \(o\) defines a transition relation
\[
T_o(\sigma,\sigma'),
\]
which specifies how a valid pre-state \(\sigma\) is transformed into a post-state \(\sigma'\) according to the EVM semantics~\cite{wood2014ethereum,hildenbrandt2018kevm}. Depending on the opcode, the transition may involve 256-bit arithmetic, stack manipulation, memory expansion, storage modification, gas accounting, or exceptional control flow.

A valid zero-knowledge proof guarantees that a witness satisfies the constraints encoded in the proving circuit. However, it does not independently guarantee that the source-level executor or witness generator faithfully implements the intended EVM semantics. Bugs such as incorrect gas charging, missing state updates, improper boundary checks, or faulty control-flow handling may therefore produce valid proofs for semantically incorrect state transitions~\cite{peng2025automated}. This work focuses on verifying such implementation-level opcode semantics before they are incorporated into the proving process.

\subsection{Formal Verification}

Formal verification establishes program correctness by expressing implementation behavior and required properties as mathematical constraints. Symbolic execution replaces concrete inputs with symbolic variables and accumulates path conditions and state-transition constraints, thereby representing multiple program executions within a single logical model~\cite{king1976symbolic}. This approach is particularly suitable for opcode implementations whose behavior depends on stack values, storage contents, gas conditions, and execution branches.

Satisfiability Modulo Theories (SMT) solvers support logical theories such as fixed-width bit-vectors, integers, arrays, and uninterpreted functions. These theories naturally model EVM word arithmetic, memory and storage structures, conditional transitions, and resource constraints~\cite{demoura2008z3,barrett2010smtlib}. Let
\(C_P(\sigma,\sigma')\) encode the behavior of an implementation \(P\), and let
\(\Phi(\sigma,\sigma')\) denote an intended semantic or safety property. Verification checks the satisfiability of
\[
C_P(\sigma,\sigma') \land \neg\Phi(\sigma,\sigma').
\]

If the formula is satisfiable, the solver returns a model that serves as a counterexample, demonstrating an execution under which the implementation violates the property. If it is unsatisfiable, the property holds for all executions captured by the encoded model and its assumptions. Although SMT solvers provide reliable reasoning backends, they require explicit, complete, and correctly typed constraints. Constructing such verification models from low-level source code therefore remains a major bottleneck. VeriSynth addresses this problem by using an LLM to synthesize executable Python/Z3 models while retaining the SMT solver as the final correctness arbiter.

\subsection{Large Language Models}

Large language models (LLMs) have demonstrated strong capabilities in source-code understanding, generation, and program repair~\cite{brown2020language,chen2021codex,hou2023llm4se}. Through pretraining on large code corpora and in-context learning, LLMs can infer relationships between variables, control-flow branches, function calls, and state updates. These capabilities make them useful for translating implementation-level program behavior into executable verification scripts.

Recent studies have explored LLMs for formal specification generation, property inference, and feedback-driven program repair~\cite{ma2025specgen,liu2024propertygpt,xia2023conversational}. Nevertheless, generating an executable verification model is more demanding than producing a natural-language explanation or a high-level property. The model must correctly represent path conditions, symbolic types, state updates, resource accounting, and solver-specific APIs. Small mistakes, such as using inconsistent bit-vector widths or omitting a branch condition, can make the generated model either non-executable or semantically unsound.

Therefore, LLM-generated artifacts cannot provide formal correctness guarantees by themselves. VeriSynth uses the LLM only as a formalization frontend that translates zkEVM source code into candidate Python/Z3 models. Compilation diagnostics, type and sort errors, solver exceptions, and counterexamples are then used as structured feedback to repair incomplete or inconsistent models. The LLM proposes the symbolic model, while the SMT solver determines whether the encoded correctness properties hold.

\section{Approach}
\label{sec:approach}

This section presents \toolname{}, an LLM-guided framework for synthesizing executable verification models from source-level implementations.
The key design principle is to use LLMs for semantic model construction while leaving correctness checking to an SMT solver.
Given an input implementation, \toolname{} decomposes the program into verification-relevant semantic units, retrieves previously validated examples as semantic anchors, synthesizes executable SMT constraints, and repairs failed models using compiler and solver feedback.

Following the generate-and-verify paradigm used in recent LLM-assisted specification generation and repair work~\cite{ma2025specgen,xia2023conversational}, \toolname{} adopts an iterative workflow in which candidate verification models are generated, checked, and refined.
However, unlike approaches that generate declarative specifications such as JML annotations, \toolname{} generates executable verification models that can be directly checked by Z3.
This distinction is important for zkEVM implementations, where execution semantics such as opcode dispatch, storage updates, gas accounting, and environment-dependent behavior are often implicit in source code rather than written as formal specifications.

\subsection{Overview and Problem Formulation}
\label{subsec:problem}

The input of \toolname{} is a source program $P$ written in a supported language.
The language-specific front end parses $P$ and normalizes it into an intermediate representation from which verification-relevant semantic units are extracted.
A semantic unit may correspond to an opcode handler, a state transition routine, a memory or storage update, or a helper function that participates in execution semantics.
\ljy{In the zkEVM context, we classify units into five semantic families: arithmetic/bitwise, memory/storage, gas/refund, call/create paths, and environment/context-dependent behavior.}

For each semantic unit $u$, \toolname{} aims to synthesize an executable verification model:
\begin{equation}
    M_u = \langle V_u, C_u, \Phi_u \rangle ,
    \label{eq:model}
\end{equation}
where $V_u$ is the set of symbolic variables, $C_u$ is the set of solver constraints encoding the behavior of $u$, and $\Phi_u$ is the property or consistency obligation to be checked.
The symbolic state of zkEVM-style code typically contains stack, memory, storage, gas, refund, warm-access information, and environment variables, following the stateful execution model of the EVM~\cite{wood2014ethereum,hildenbrandt2018kevm}. The generated constraints describe how such a state is transformed by the target unit.

The verification task is to check whether the synthesized model admits a counterexample under the chosen abstraction.
\ljy{Given the model in Eq.~\ref{eq:model},} if $C_u \land \neg \Phi_u$ is unsatisfiable, the model is accepted with respect to $\Phi_u$; otherwise, the solver-produced counterexample is treated as feedback for repair, indicating a potential semantic inconsistency in the original implementation.

This formulation makes the goal of \toolname{} different from test generation or natural-language explanation. The framework does not ask the LLM to decide whether a program is correct. Instead, the LLM proposes an executable model, and the solver validates whether the proposed model satisfies the intended constraints.

VeriSynth separates the model extracted from the target implementation from the obligations used to evaluate it. For a semantic unit \(u\), \(C_u^{\mathit{impl}}(\sigma,\sigma')\) encodes the behavior of the source implementation, while \(\Phi_u\) is constructed from validated opcode semantics and family-specific consistency rules rather than from the target code alone. The verification query is
\begin{equation}
C_u^{\mathit{impl}}(\sigma,\sigma') \land \neg\Phi_u(\sigma,\sigma').
\end{equation}
Typical obligations include arithmetic equivalence, stack-height preservation, storage-update consistency, gas and refund bounds, and correct failure or return-value propagation. A satisfiable query yields an input state under which the implementation violates the expected behavior. This separation prevents an incorrect but internally consistent implementation model from being accepted solely because the generated constraints reproduce the same faulty logic.

\subsection{Semantic Decomposition and Retrieval}
\label{subsec:decomposition}

\ljy{The formulation above assumes a semantic unit $u$ is already isolated. In practice, raw zkEVM source files intermix opcode handlers, helper routines, and state objects.} Directly applying LLMs to such files is unreliable because the relevant semantics are scattered across these components.
Therefore, \toolname{} first decomposes the input into semantically coherent units.
The purpose of decomposition is not merely to reduce prompt length, but to expose state effects that can be modeled independently while preserving explicit dependencies through unit interfaces.

\begin{figure*}[t]
  \centering
  \includegraphics[width=\textwidth]{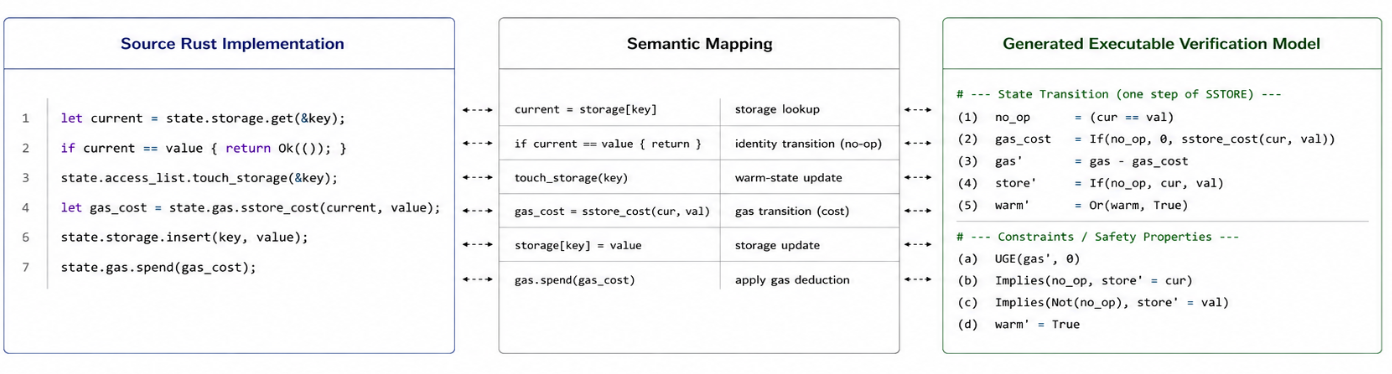}
  \caption{Example of semantic-to-SMT synthesis. Given a Rust implementation of the SSTORE opcode, VeriSynth maps key semantic operations to precise state-transition constraints and safety properties, producing an executable verification model ready for SMT solving.}
  \label{fig:code}
\end{figure*}

\ljy{For each code fragment, VeriSynth identifies which state components it reads and writes, and assigns it to a semantic family. Fragments contributing to the same state transition are grouped into one unit.}
For example, an SSTORE-like unit may include reading the current storage value, detecting the no-op case, updating the warm-access state, computing gas cost, and writing the new storage value.
Although these operations appear as separate statements in Rust, they jointly define one semantic transition and should be modeled together. \ljy{Figure~\ref{fig:code} illustrates this mapping for an SSTORE unit.}

Retrieval-augmented prompting has been used to provide LLMs with task-specific context in code and specification generation tasks~\cite{liu2024propertygpt,white2023prompt}.
In \toolname{}, retrieval is used differently: the retrieved examples serve as semantic anchors for executable model construction. After decomposition, \toolname{} retrieves similar verified examples from a reference database:
\[
    \mathcal{D} = \{(u^r, M^r, meta^r)\},
\]
where $u^r$ is a reference unit, $M^r$ is its verified executable model, and $meta^r$ records metadata such as opcode family, state components, and modeling patterns.

Given a target unit $u$ and a reference unit $d$, their similarity is computed as:
\begin{equation}
\begin{aligned}
    \textsc{Sim}(u,d) =
    \alpha \cdot
    \frac{\textsc{Emb}(u)\cdot \textsc{Emb}(d)}
    {\|\textsc{Emb}(u)\|\|\textsc{Emb}(d)\|}
    +
    (1-\alpha)\cdot \textsc{Fam}(u,d),
\end{aligned}
\label{eq:similarity}
\end{equation}
where $\textsc{Emb}(\cdot)$ denotes the code embedding and $\textsc{Fam}(u,d)$ rewards matches in opcode family or semantic category.
\ljy{Based on this similarity measure (Eq.~\ref{eq:similarity}),} the top-$k$ examples are used as semantic anchors for synthesis.

\ljy{The retrieved examples are not copied as templates. Instead, they serve as a modeling prior: they stabilize the symbolic vocabulary, state schema, and transition structure by grounding generation in previously validated modeling decisions.}
Each reference entry is added to the retrieval database only after its
executable model passes syntax checking, sort checking, solver execution,
and the associated semantic obligations. During retrieval, VeriSynth
prioritizes examples with compatible opcode families, accessed state
components, and symbolic types. The retrieved models guide symbolic
declarations, transition structures, and solver usage rather than being
directly copied. To prevent evaluation leakage, examples sharing the
same source-file identity or opcode unit with the target are excluded
from the candidate set.

\subsection{LLM-Guided Constraint Synthesis}
\label{subsec:synthesis}

VeriSynth uses a typed symbolic state shared across generated models.
EVM words are represented as fixed-width bit-vectors, control flags as
Boolean expressions, and memory and storage as symbolic arrays. The
state is divided into pre-state and post-state variables:
\begin{equation}
\begin{aligned}
\sigma=\langle stk,mem,sto,gas,ref,ctx,status\rangle, \\
\sigma'=\langle stk',mem',sto',gas',ref',ctx',status'\rangle.
\end{aligned}
\end{equation}
Source-level branches are translated into guarded constraints or Z3
\texttt{If} expressions, while Rust assignments define equalities over
post-state variables. Each state component must either be explicitly
updated or connected to its pre-state through a frame condition. For
example, an opcode that does not modify memory introduces
\(mem'=mem\). These frame conditions prevent omitted updates from
leaving post-state variables unconstrained. The encoding also enforces
consistent bit-vector widths and explicit status variables for
exceptional exits, reducing ambiguity during constraint synthesis and
providing structured targets for solver-guided repair.

LLMs have shown strong capabilities in code generation and code understanding~\cite{chen2021codex,openai2023gpt4}.
In \toolname{}, we use this capability to translate each semantic unit into an executable SMT model rather than a natural-language description.
For a target unit $u$, the prompt contains four parts: the source code of $u$, the retrieved semantic anchors, opcode-aware modeling instructions, and the required output format.
\ljy{The opcode-aware instructions specify the relevant symbolic state schema (stack, memory, storage, gas, environment variables) and modeling obligations for the semantic family of $u$, guiding the LLM to encode the appropriate variable types, transition structure, and safety assertions.}
When repair is needed, compiler diagnostics or solver feedback are additionally appended to the prompt.

\ljy{The synthesized constraint set falls into four categories: guard and path conditions, state-transition constraints, resource constraints such as gas and refund behavior, and assertions or safety obligations.}

\ljy{To illustrate the intended mapping from code to constraints, consider a simplified SSTORE unit (illustrative example):}
\[
\begin{aligned}
    no\_op &:= (cur = val), \\
    warm' &:= true, \\
    gas\_cost &:= ite(no\_op,0,sstore\_cost(cur,val)), \\
    gas' &:= gas - gas\_cost, \\
    store' &:= ite(no\_op,cur,val).
\end{aligned}
\]
The corresponding safety obligations can be written as:
\[
\begin{aligned}
    gas' &\geq 0, \\
    no\_op &\Rightarrow store' = cur, \\
    \neg no\_op &\Rightarrow store' = val.
\end{aligned}
\]

\subsection{Executable Verification and Auto-Repair}
\label{subsec:repair}

The generated model may still fail verification.
Common failures include missing declarations, invalid Python or Z3 API usage, bit-vector sort mismatches, incomplete state updates, or semantic inconsistencies exposed by counterexamples.
Instead of discarding failed models, \toolname{} repairs them using structured feedback from the compiler and solver, following the broader idea of feedback-driven LLM repair~\cite{xia2023conversational}.

\ljy{As formalized in Eq.~\ref{eq:verify}, given} a candidate model $\widehat{M}_u$, the verifier returns:
\begin{equation}
    \textsc{Verify}(\widehat{M}_u)=\langle ok,e,w\rangle,
    \label{eq:verify}
\end{equation}
where $ok$ indicates whether the model passes verification, $e$ is the error class, and $w$ is an optional witness such as a compiler trace or solver counterexample.

Verification proceeds in four stages. VeriSynth first parses and imports
the generated Python script, then constructs the symbolic expressions to
detect missing declarations and sort mismatches. Next, it executes the
solver query under a bounded timeout and finally checks the semantic
obligations. When a stage fails, only the corresponding diagnostic and
the relevant model fragment are returned to the LLM, enabling localized
repair while preserving constraints that have already passed earlier
checks.

We classify repair-relevant failures into four categories:
\[
    \mathcal{E} =
    \{E_{\text{parse}}, E_{\text{sort}}, E_{\text{exec}}, E_{\text{sem}}\}.
\]
Here, $E_{\text{parse}}$ denotes syntax or missing-declaration errors, $E_{\text{sort}}$ denotes type or sort mismatches, $E_{\text{exec}}$ denotes solver execution failures or timeouts, and $E_{\text{sem}}$ denotes semantic inconsistencies exposed by solver witnesses.

For each error class $e$, \toolname{} maintains a candidate repair set $\mathcal{R}(e)$.
A repair action may complete missing declarations, correct bit-vector widths, replace invalid operators, refine guard constraints, patch state updates, or regenerate a localized fragment of the model.
\ljy{Repair actions are applied in a fixed priority per error class (Table~\ref{tab:repair-actions}), favoring actions with minimal structural modification.}

Table~\ref{tab:algorithm-verisynth} summarizes the overall procedure.
The loop terminates when the model passes all verification phases or the repair budget is exhausted.
This bounded process prevents unbounded regeneration and makes the system behavior measurable.

\begin{table}[t]
\centering
\caption{LLM-guided synthesis and repair.}
\label{tab:algorithm-verisynth}
\small
\begin{tabularx}{\columnwidth}{p{0.18\columnwidth}X}
\toprule
\textbf{Input} &
\ljy{$P$ (source program), $\mathcal{D}$ (reference database), $\mathcal{R}$ (repair action set), $k$ (retrieval count), $T_{\max}$ (repair budget).} \\

\midrule
\textbf{Output} &
\ljy{Verified executable models for all accepted units.} \\

\midrule
\textbf{Procedure} &
\begin{minipage}[t]{0.76\columnwidth}
\vspace{-0.2em}
\begin{enumerate}[leftmargin=*, itemsep=0pt, topsep=0pt]
    \item Parse and normalize $P$.
    \item Decompose $P$ into semantic units.
    \item Initialize $M_{\text{final}} \leftarrow \emptyset$.
    \item For each unit $u$:
    \begin{enumerate}[leftmargin=*, itemsep=0pt, topsep=0pt]
        \item Identify semantic family $f(u)$.
        \item Retrieve top-$k$ anchors from $\mathcal{D}$.
        \item Synthesize candidate model $\widehat{M}_u$.
        \item Verify $\widehat{M}_u$.
        \item If verification fails, \ljy{repair the model using compiler/solver feedback.}
        \item Repeat until accepted or $T_{\max}$ is reached.
    \end{enumerate}
\end{enumerate}
\vspace{-0.2em}
\end{minipage}
\\
\bottomrule
\end{tabularx}
\end{table}

\begin{table}[t]
\centering
\caption{Error categories and repair actions in \toolname{}.}
\label{tab:repair-actions}
\scriptsize
\renewcommand{\arraystretch}{1.12}
\begin{tabular}{|p{0.10\columnwidth}|p{0.18\columnwidth}|p{0.53\columnwidth}|}
\hline
\textbf{Error} & \textbf{Source} & \textbf{Repair Action} \\
\hline
Parse & Compiler & Fix syntax, imports, and missing declarations. \\
\hline
Sort & Z3 exception & Align bit-vector widths and symbolic sorts. \\
\hline
Execution & Traceback & Repair invalid APIs and solver construction. \\
\hline
Semantic & Counterexample & Refine guards, transitions, and constraints. \\
\hline
\end{tabular}
\end{table}

Finally, \toolname{} is verification-guided rather than LLM-trusting.
The LLM proposes candidate models; \ljy{the compiler and SMT solver check their executability and internal consistency. This shifts error detection from LLM judgment to machine-checkable consistency.}
Thus, the framework uses LLMs to reduce the manual burden of formal modeling while retaining solver-backed checking as the final validation mechanism.

The guarantees of VeriSynth are relative to the extracted semantic unit,
symbolic abstraction, and encoded obligations. The LLM is not treated as
a correctness oracle; its output must pass syntax, sort, execution, and
semantic checks. Thus, an unsatisfiable query establishes correctness
only within the modeled state and assumptions, rather than for the entire
zkEVM implementation.

\section{Experimental Setup}
\label{sec:setup}

In this section, we introduce the experimental design, dataset, baselines, and evaluation metrics used to evaluate \toolname{}.
Our experiments are organized around the following research questions:

\begin{itemize}
    \item \textbf{RQ1: Effectiveness.} How accurately can \toolname{} detect opcode-related semantic bugs?
    \item \textbf{RQ2: Opcode-Family Generalization.} Does \toolname{} remain effective across different opcode semantic families?
    \item \textbf{RQ3: Ablation.} How much do semantic decomposition, retrieval grounding, opcode-aware prompting, auto-repair, and solver-backed validation contribute to the final performance?
    \item \textbf{RQ4: Token Cost.} What is the token cost of the full verification pipeline?
    \item \textbf{RQ5: Practical Value.} Compared with Scroll's handwritten Rust mutation-test workflow, what additional bug-detection value does \toolname{} provide?
\end{itemize}

\subsection{Implementation}
\label{subsec:implementation}

We implement \toolname{} as a prototype framework that integrates language-specific front ends, retrieval-guided prompting, executable Python/Z3 model synthesis, and verification-guided repair.
\ljy{The current prototype supports source-level analysis of Rust programs and generates Python/Z3 verification scripts.}
For each target program, \toolname{} first extracts verification-relevant semantic units, retrieves similar verified examples from a reference database, generates executable SMT models, and invokes Z3 to check the synthesized constraints.

We use GPT-4o as the backend LLM in the current evaluation.
The maximum number of repair iterations is set to 3.
For each sample, we record the final detection result, executability status, repair behavior, and token usage.
Since the current prototype is intended for offline verification of security-critical implementation logic, our cost analysis focuses on token consumption rather than online runtime overhead.

\subsection{Dataset}
\label{subsec:datasets}

We evaluate \toolname{} on a self-constructed benchmark of 95 negative samples derived from zkEVM-style opcode implementation logic.
Each sample contains one localized opcode-related bug injected into a target semantic unit while preserving most of the surrounding code structure.
The benchmark is designed to evaluate whether \toolname{} can detect semantic inconsistencies in implementation-level opcode logic.

\subsubsection{Negative-sample Construction}
To construct the benchmark, we inject one bug into each target unit.
The injected errors include wrong arithmetic rules, incorrect boundary handling, missing state updates, wrong return-value propagation, incorrect gas/refund logic, and execution-path mistakes in handlers such as \texttt{CALL} and \texttt{CREATE}.
The samples cover five opcode semantic families: arithmetic and bitwise operations, memory and storage operations, gas and refund logic, call/create paths, and environment/context-dependent behavior.

To avoid retrieval leakage, reference examples used by the retrieval module are isolated from target samples by source-file identity and opcode unit.
Thus, the retrieved examples provide semantic anchors for modeling without directly exposing the target mutation.

\subsubsection{Positive Samples}
\ljy{In addition to negative samples, we use original non-mutated implementation units for internal sanity checking and executability analysis. The main reported effectiveness results focus on the 95 negative samples.}

\subsection{Baselines}
\label{subsec:baselines}

We compare \toolname{} with the following baselines.

\textbf{LLM-only.}
This baseline directly asks the LLM to identify the bug or generate a verification script without semantic decomposition, retrieved examples, opcode-aware prompts, Z3-backed validation, or structured auto-repair.
It represents the simplest LLM-based setting.

\textbf{Conversational LLM.}
This baseline allows the model to revise its output using compiler or solver feedback.
However, it does not use retrieved semantic anchors or structured repair actions.
This baseline is used to evaluate whether feedback alone is sufficient for robust verification model construction.

\textbf{Handwritten Rust tests.}
For practical comparison, we use Scroll's handwritten Rust mutation-test workflow as an engineering baseline.
For each sample, the mapped Rust source file is patched, the corresponding \texttt{cargo test} entry is executed, and the original implementation is restored after the test.
\ljy{We count both assertion failures and compilation failures as successful exposures, since both prevent the mutated implementation from passing the test workflow.}
This baseline is used in RQ5.

\textbf{\toolname{}.}
The full system includes semantic decomposition, retrieval-guided constraint synthesis, opcode-aware prompting, executable Z3 verification, and auto-repair.

\subsection{Evaluation Metrics}
\label{subsec:metrics}

We use the following metrics to evaluate effectiveness and cost.

\textbf{Detection Rate (DR).}
DR measures the fraction of negative samples correctly identified as semantically inconsistent:
\[
    DR = \frac{N_{\text{detected}}}{N_{\text{negative}}}.
\]
\ljy{A bug is considered detected when the synthesized verification model is executable and the SMT solver finds a counterexample ($C_u \land \neg \Phi_u$ satisfiable), indicating a semantic inconsistency between the encoded constraints and the safety obligations. Detection rate is reported on negative samples only; systematic false-alarm estimation on positive samples is deferred to future work.}

\textbf{Executability Rate (ER).}
ER measures the fraction of generated verification models that can be successfully compiled and executed:
\[
    ER = \frac{N_{\text{executable}}}{N_{\text{total}}}.
\]


\textbf{Token Cost.}
We report the average token usage per sample and estimate the total token consumption for one full benchmark run.
This metric reflects the cost of LLM-based synthesis and repair.

\section{Experimental Results}
\label{sec:experiments}

This section reports the experimental results of \toolname{}.
We organize the results according to the five research questions introduced in Section~\ref{sec:setup}.
Overall, the results show that \toolname{} substantially outperforms LLM-only baselines in opcode-level bug detection, remains effective across different opcode semantic families, and benefits from each major component in the pipeline.
We further analyze the token cost of the framework and compare its practical value with Scroll's handwritten Rust mutation-test workflow.

\subsection{RQ1: Overall Bug Detection Effectiveness}
\label{subsec:rq1}

RQ1 evaluates whether \toolname{} can effectively detect injected opcode-level semantic bugs.
We conduct the experiment on the 95-sample negative benchmark.
Each sample contains one localized opcode-related bug, such as an incorrect arithmetic rule, missing state update, wrong boundary condition, incorrect gas/refund logic, or execution-path error in handlers such as \texttt{CALL} and \texttt{CREATE}.

\ljy{We compare \toolname{} with the baselines described in Section~4.3: LLM-only, Conversational LLM, and Handwritten Rust tests (for RQ5).}

\begin{table}[t]
\centering
\caption{Overall bug detection effectiveness on the 95-sample benchmark.}
\label{tab:rq1-effectiveness}
\small
\renewcommand{\arraystretch}{1.12}
\begin{tabular}{|l|c|c|}
\hline
\textbf{Approach} & \textbf{Detected / Total} & \textbf{DR} \\
\hline
LLM-only & 44 / 95 & 46.3\% \\
\hline
Conversational LLM & 58 / 95 & 61.1\% \\
\hline
\toolname{} & 87 / 95 & 91.6\% \\
\hline
\end{tabular}
\end{table}
The conversational baseline is inspired by feedback-driven LLM repair settings~\cite{xia2023conversational}.
As shown in Table~\ref{tab:rq1-effectiveness}, \toolname{} detects 87 out of 95 injected bugs, achieving a detection rate of 91.6\% (DR defined in Section~4.4).
In contrast, the LLM-only baseline detects only 44 bugs, corresponding to a detection rate of 46.3\%.
Although the conversational baseline improves over LLM-only generation by using feedback, it still detects only 58 bugs.
These results indicate that feedback alone is insufficient for reliable verification model construction.
The improvement of \toolname{} comes from the combination of semantic decomposition, retrieval-grounded modeling, opcode-aware prompts, and solver-guided repair.

The result also supports our central design choice: LLMs should not be used as final bug judges.
Instead, they are more effective when used as semantic model constructors, while the final validation is performed by an SMT solver.
\ljy{This distinction explains why \toolname{} outperforms direct LLM-based detection.}

\noindent\fbox{
\begin{minipage}{0.95\linewidth}
\textbf{RQ1:}
\toolname{} detects 87 out of 95 injected opcode-level bugs, achieving a detection rate of 91.6\%.
This substantially outperforms LLM-only detection and conversational LLM repair, showing the effectiveness of solver-backed verification model synthesis.
\end{minipage}
}

\subsection{RQ2: Effectiveness Across Opcode Semantic Families}
\label{subsec:rq2}

RQ2 investigates whether \toolname{} remains effective across different opcode semantic families.
This question is important because zkEVM implementation bugs are not limited to simple arithmetic operations.
Many difficult cases involve storage updates, memory behavior, call/create execution paths, gas accounting, and environment-dependent logic.
Therefore, we divide the 95 negative samples into five semantic families and report the detection results for each family.

\begin{table}[t]
\centering
\caption{Bug detection results across opcode semantic families.}
\label{tab:rq2-family}
\small
\renewcommand{\arraystretch}{1.12}
\begin{tabular}{|l|c|c|}
\hline
\textbf{Opcode Family} & \textbf{Detected / Total} & \textbf{DR} \\
\hline
Arithmetic and Bitwise & 21 / 22 & 95.5\% \\
\hline
Memory and Storage & 24 / 26 & 92.3\% \\
\hline
Gas and Refund & 14 / 15 & 93.3\% \\
\hline
Call/Create Paths & 18 / 21 & 85.7\% \\
\hline
Environment and Context & 10 / 11 & 90.9\% \\
\hline
Overall & 87 / 95 & 91.6\% \\
\hline
\end{tabular}
\end{table}

Table~\ref{tab:rq2-family} shows that \toolname{} performs well across all semantic families.
The highest detection rate is achieved on arithmetic and bitwise operations, where the state transition is relatively local and can be naturally encoded using bit-vector constraints.
The framework also performs strongly on memory/storage and gas/refund cases, demonstrating that it can model nontrivial state changes beyond simple numerical operations.

The most challenging category is call/create paths, where \toolname{} detects 18 out of 21 bugs.
This lower detection rate is expected because call/create semantics involve multiple interacting components, including execution context, return values, address creation, memory effects, and failure propagation.
Even in this category, \toolname{} still achieves an 85.7\% detection rate, suggesting that the proposed semantic decomposition and retrieval grounding are useful for complex opcode-level behaviors.

\noindent\fbox{
\begin{minipage}{0.95\linewidth}
\textbf{RQ2:}
\toolname{} remains effective across different opcode semantic families.
While call/create paths are more challenging than arithmetic operations, the framework still detects most injected bugs in complex state-transition scenarios.
\end{minipage}
}

\subsection{RQ3: Ablation Study}
\label{subsec:rq3}

RQ3 evaluates the contribution of each major component in \toolname{}.
We consider five ablated variants.
The first removes semantic decomposition and directly feeds larger code fragments to the LLM.
The second removes retrieval grounding and generates verification models without semantically similar reference examples.
The third removes opcode-aware prompting and uses a generic prompt for all opcode families.
The fourth removes auto-repair and accepts only the initially generated verification model.
The fifth removes Z3 validation, reducing the system to an LLM-only judgment setting.

\begin{table}[t]
\centering
\caption{Ablation study of \toolname{} components on the 95-sample benchmark.}
\label{tab:rq3-ablation}
\small
\renewcommand{\arraystretch}{1.12}
\begin{tabular}{|l|c|c|c|}
\hline
\textbf{Variant} & \textbf{Detected / Total} & \textbf{DR} & \textbf{ER} \\
\hline
w/o Decomposition & 72 / 95 & 75.8\% & 81.1\% \\
\hline
w/o Retrieval & 68 / 95 & 71.6\% & 78.9\% \\
\hline
w/o Opcode-aware Prompt & 74 / 95 & 77.9\% & 84.2\% \\
\hline
w/o Auto-Repair & 63 / 95 & 66.3\% & 66.3\% \\
\hline
w/o Z3 Validation & 45 / 95 & 47.4\% & -- \\
\hline
Full \toolname{} & 87 / 95 & 91.6\% & 93.7\% \\
\hline
\end{tabular}
\end{table}

Table~\ref{tab:rq3-ablation} shows that removing any major component reduces the final performance.
The largest drop occurs when auto-repair is removed.
Without auto-repair, the detection rate decreases from 91.6\% to 66.3\%, and the executability rate also drops to 66.3\%.
This indicates that many initially generated Python/Z3 scripts are semantically close to the desired model but fail due to local syntax errors, missing declarations, invalid solver API usage, or bit-vector sort mismatches.
The repair loop is therefore essential for turning near-valid candidate models into executable verification artifacts.
\ljy{Qualitatively, most repair actions addressed parse and sort errors (missing declarations, bit-vector width mismatches), while fewer involved semantic counterexample-guided refinement.}

Removing retrieval grounding also causes a clear performance drop, reducing the detection rate to 71.6\%.
Without retrieved examples, the LLM is more likely to produce inconsistent symbolic variables, incomplete transition relations, or missing resource constraints.
This confirms that retrieval acts as a modeling prior rather than ordinary context augmentation.

Removing semantic decomposition reduces the detection rate to 75.8\%.
This variant particularly struggles with complex opcode handlers, where relevant state effects are scattered across helper routines, state objects, and conditional branches.
By contrast, semantic decomposition isolates verification-relevant units and makes the model synthesis task more tractable.

Removing opcode-aware prompting reduces the detection rate to 77.9\%.
This result suggests that generic prompts are insufficient for diverse opcode families.
Arithmetic operations, storage updates, gas accounting, and call/create paths require different symbolic state schemas and modeling obligations.

Finally, the variant without Z3 validation detects only 45 out of 95 bugs.
This result is close to the LLM-only baseline and confirms that the final judgment must remain solver-backed.
The LLM can propose candidate models, but Z3 is necessary for reliable verification.

\noindent\fbox{
\begin{minipage}{0.95\linewidth}
\textbf{RQ3:}
All major components contribute to the effectiveness of \toolname{}.
Auto-repair has the largest impact on executability, while retrieval and decomposition improve semantic consistency and robustness on complex opcode handlers.
\end{minipage}
}

\subsection{RQ4: Token Cost Analysis}
\label{subsec:rq4}

RQ4 analyzes the token cost of \toolname{}.
Since the current prototype relies on LLM calls for constraint synthesis and repair, token usage is an important practical factor.
We therefore measure the average token consumption per sample and estimate the total cost for one full run over the 95-sample benchmark.

\begin{table}[t]
\centering
\caption{Token cost of \toolname{} on the 95-sample benchmark.}
\label{tab:rq4-token-cost}
\small
\renewcommand{\arraystretch}{1.12}
\begin{tabular}{|l|c|}
\hline
\textbf{Metric} & \textbf{Value} \\
\hline
Benchmark size & 95 samples \\
\hline
Backend model & GPT-4o \\
\hline
Average tokens per sample & $\sim$100K \\
\hline
Estimated total tokens & $\sim$9.5M \\
\hline
Average repair rounds & 1.8 \\
\hline
Maximum repair budget & 3 \\
\hline
\end{tabular}
\end{table}

As shown in Table~\ref{tab:rq4-token-cost}, each sample consumes approximately 100K tokens on average.
This includes the target semantic unit, retrieved anchors, opcode-aware modeling instructions, generated Python/Z3 scripts, compiler diagnostics, solver feedback, and repair prompts.
For the 95-sample benchmark, the estimated total token usage is approximately 9.5M tokens.

The token cost is nontrivial.
However, it is bounded by the repair budget and predictable in practice.
In our current setting, the maximum number of repair iterations is limited to 3, and the average number of repair rounds is 1.8.
This suggests that most failed generations can be repaired within a small number of iterations.

It is also important to note that \toolname{} is designed for offline verification of security-critical implementation logic rather than online per-transaction execution.
Therefore, the token overhead is acceptable for scenarios such as opcode implementation review, regression checking after code changes, and validation of high-risk state-transition logic.
Compared with the manual effort required to construct Python/Z3 verification models, the LLM cost provides a practical tradeoff between automation and verification rigor.

\noindent\fbox{
\begin{minipage}{0.95\linewidth}
\textbf{RQ4:}
Each sample consumes approximately 100K tokens on average, resulting in about 9.5M tokens for the 95-sample benchmark.
Although the cost is nontrivial, it remains bounded by the repair budget and is suitable for offline verification of security-critical zkEVM implementation logic.
\end{minipage}
}

\subsection{RQ5: Practical Value Compared with Handwritten Rust Tests}
\label{subsec:rq5}

RQ5 evaluates whether \toolname{} provides practical value beyond a handwritten
Rust mutation-test workflow.
\ljy{For this comparison, we report results grouped into two high-level categories that complement the per-family breakdown in RQ2: opcode semantic errors (spanning arithmetic, memory/storage, gas/refund, and environment families) and opcode execution-path errors (spanning call/create path and branch logic).}
For each negative sample, the handwritten
baseline temporarily patches the mapped Scroll source file, runs the
corresponding \texttt{cargo test} entry in the original crate environment, and
then restores the original implementation. This baseline reflects a realistic
engineering workflow: it uses tests that are already maintained with the
project, but its effectiveness is bounded by the assertions and execution cases
that developers have manually written.

\begin{table}[t]
\centering
\caption{Practical comparison with Scroll's handwritten Rust tests.}
\label{tab:rq5-practical}
\small
\renewcommand{\arraystretch}{1.12}
\begin{tabularx}{\columnwidth}{|X|c|c|c|}
\hline
\textbf{Bug Group} & \textbf{Total} & \textbf{Handwritten} & \textbf{\toolname{}} \\
\hline
Opcode semantic errors & 75 & 44 & 69 \\
\hline
Opcode execution-path errors & 20 & 11 & 18 \\
\hline
Overall & 95 & 55 & 87 \\
\hline
\end{tabularx}
\end{table}

\begin{figure}[t]
    \centering
    \includegraphics[width=\linewidth]{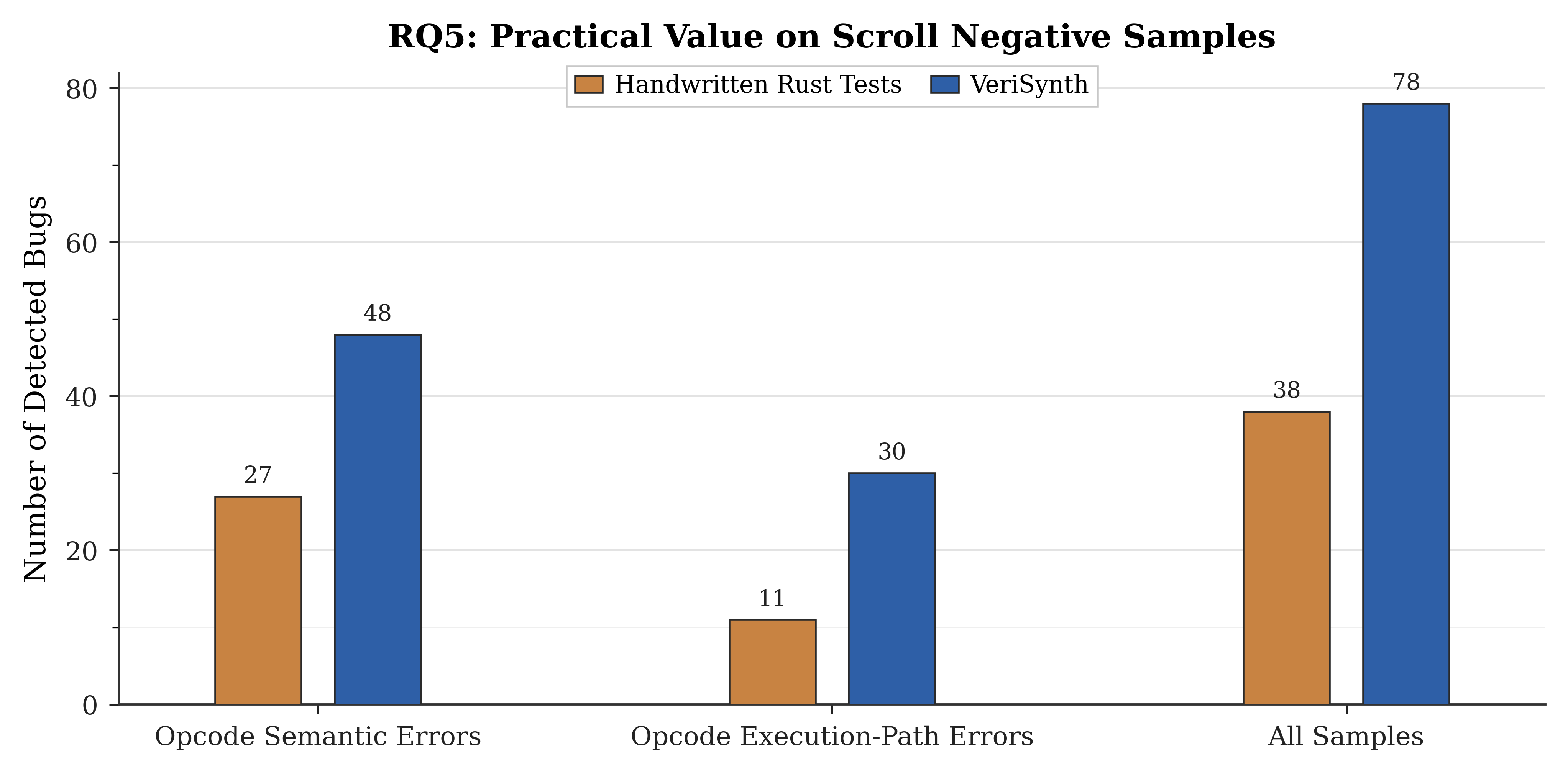}
    \caption{Practical comparison between handwritten Rust tests and \toolname{} \ljy{by bug category} on the \texttt{test\_neg} benchmark.}
    \label{fig:rq5-practical-value}
\end{figure}

Table~\ref{tab:rq5-practical} and Fig.~\ref{fig:rq5-practical-value} report
the comparison on our 95-sample negative benchmark, denoted as
\texttt{test\_neg}. The handwritten Rust workflow exposes 55 injected bugs,
while \toolname{} detects 87. The gap is consistent across both groups: for
opcode semantic errors, handwritten tests expose 44 out of 75 cases, whereas
\toolname{} detects 69; for opcode execution-path errors, handwritten tests
expose 11 out of 20 cases, whereas \toolname{} detects 18. Overall, this
corresponds to a detection rate of 57.9\% for handwritten tests and 91.6\% for
\toolname{}.

Even under this \ljy{counting (see Section~4.3)}, handwritten tests leave a substantial fraction of
negative samples unexposed. \toolname{} improves coverage by synthesizing
explicit semantic constraints and checking them with Z3, which makes it less
dependent on whether a particular behavior has already been encoded as a
manual unit-test assertion.

\noindent\fbox{
\begin{minipage}{0.95\linewidth}
\textbf{RQ5:}
Compared with Scroll's handwritten Rust mutation-test workflow, \toolname{}
detects 87 bugs on \texttt{test\_neg}, while handwritten tests expose 55.
The improvement across both semantic and execution-path bug groups indicates
that \toolname{} is a practical complement to manually maintained Rust tests.
\end{minipage}
}

\section{Related Work}
\label{sec:related-work}

\subsection{Formal Semantics and Verification of EVMs}

A substantial body of work has studied the formal semantics and verification of Ethereum programs. KEVM provides an executable formal semantics of the EVM in the K framework and enables reasoning about bytecode-level execution~\cite{hildenbrandt2018kevm}. Tools such as ZEUS and Securify analyze smart contracts against predefined safety properties using abstract interpretation, symbolic reasoning, or compliance and violation patterns~\cite{kalra2018zeus,tsankov2018securify}. These approaches primarily target smart contracts or EVM bytecode and generally assume that the underlying virtual-machine implementation correctly realizes the intended semantics.

Recent work has also examined the soundness and completeness of zkEVM constraint systems. For example, automated vetting techniques detect under-constrained or over-constrained behaviors in zkEVM circuits and witness generation~\cite{peng2025automated}. Such work is complementary to VeriSynth: circuit-level verification checks whether a proving system correctly constrains its witnesses, whereas VeriSynth focuses on whether source-level Rust opcode handlers correctly implement the intended state transitions. In addition, existing formal-verification approaches usually require manually written semantics or properties. VeriSynth instead synthesizes executable Python/Z3 models directly from source-level implementations.

\subsection{Testing and Validation of EVM Implementations}

Production EVM and zkEVM projects rely extensively on unit tests, integration tests, official Ethereum execution vectors, and regression testing. These techniques are effective for checking known corner cases and preventing previously discovered defects from reappearing. Mutation testing can further evaluate whether a test suite exposes deliberately injected implementation faults.

Differential testing provides another practical validation strategy. EVMFuzz, for example, generates and mutates contract inputs and compares the resulting executions across different EVM implementations~\cite{fu2024evmfuzz}. Disagreements in return values, traces, or gas consumption can reveal implementation defects. However, testing and fuzzing reason about a finite set of concrete executions, and their effectiveness depends on the selected inputs, assertions, and comparison oracles. Bugs involving uncommon storage states, gas boundaries, exceptional branches, or interactions between execution components may therefore remain unexposed. VeriSynth complements these techniques by constructing symbolic transition models and using SMT-generated counterexamples to reason about classes of executions rather than individual test inputs.

\subsection{Large Language Models for Software Engineering}

Large language models (LLMs) have been increasingly applied to software engineering tasks, including code understanding, code generation, test generation, program repair, and code summarization~\cite{brown2020language,chen2021codex,openai2023gpt4,hou2023llm4se}. Their in-context learning ability allows developers to provide task descriptions, demonstrations, and feedback through prompts~\cite{brown2020language,white2023prompt}. Instruction tuning and reasoning-oriented prompting can further improve the controllability and consistency of generated artifacts~\cite{ouyang2022instructgpt,wei2022chain,wang2023selfconsistency}. Recent studies have extended these capabilities to formal specification generation. SpecGen generates program specifications from source code, while PropertyGPT combines retrieval with LLM generation to infer properties for smart contracts~\cite{ma2025specgen,liu2024propertygpt}. These approaches mainly produce declarative properties or annotations, whereas zkEVM verification requires executable models that encode complete state transitions across multiple execution-state components.

LLMs have also been used to improve software artifacts through execution feedback. Conversational automated program repair, for example, iteratively revises candidate patches using compiler errors, test failures, and runtime diagnostics~\cite{xia2023conversational}. However, LLM-generated verification models may contain hallucinated semantics, omitted path conditions, inconsistent symbolic types, or invalid solver operations. Consequently, our goal is not to use an LLM as a direct bug detector or correctness oracle. Instead, \toolname{} uses the LLM as a formalization frontend that translates implementation-level opcode semantics into executable SMT-based verification models. Semantic decomposition and retrieval-grounded prompting guide model generation, while compiler diagnostics and solver feedback support bounded auto-repair. The final correctness decision is always made by the SMT solver.

\section{Conclusion}
\label{sec:conclusion}

We introduced VeriSynth, an LLM-guided framework that synthesizes executable Python/Z3 verification models from zkEVM Rust source code by combining semantic decomposition, retrieval-grounded prompting, and verification-guided auto-repair.
On our self-constructed benchmark of 95 injected opcode-level bugs, VeriSynth detects 91.6\%, outperforming direct LLM baselines (46.3--61.1\%) and Scroll's handwritten mutation-test suite (57.9\%).

Our framework remains effective across all five opcode semantic families, including complex call/create paths (85.7\%).
Ablation confirms that auto-repair contributes the largest gain in executability, followed by retrieval grounding.
Compared with Scroll's handwritten tests, VeriSynth detects 32 additional bugs, indicating that automated constraint synthesis complements manual testing.

Future work includes systematic false-alarm estimation on positive samples, cross-project validation on additional zkEVM implementations, and cost reduction through smaller model distillation.

\bibliographystyle{IEEEtran}
\bibliography{references}

\end{document}